# Use of time-correlated single photon counting detection to measure the speed of light in water*

by

Pedro L. Muiño[a], Aaron M. Thompson[b], Robert J. Buenker[c]

[a]Department of Chemistry, Mathematics and Physical Sciences, St. Francis University, Loretto, Pennsylvania 15940, USA

[b]Department of Chemistry, Kansas State University, Manhattan, Kansas 66506 USA

[c]Bergische Universität Wuppertal, Fachbereich C-Theoretische Chemie, Gaussstr. 20, D-42097 Wuppertal, Germany

**Abstract**

Traditional methods for measuring the speed of light in dispersive media have been based on the detection of interference between light waves emitted from the same source. In the present study the elapsed times for single photons to move from a laser to a photomultiplier tube are measured electronically. Time-correlated single photon counting detection produces a characteristic instrument response which has the same shape independent of both the path length the light travels and the nature of the transparent media through which it passes. This allows for an accurate calibration of the chronograph by observing shifts in the location of the instrument response for different distances traveled by the light. Measurement of the corresponding shift which occurs when light moves the same distance through air and water then enables an accurate determination of the ratio of the photon velocities in these two media. Three different wavelengths of light have been used. In two cases good agreement is found between the present measured light speeds and those which can be inferred from existing refractive index measurements in water. The shortest wavelength studied is too far in the uv to obtain a reliable estimate on the same basis, and so the $n_g$ value (1.463) measured in the present work awaits independent confirmation. A theoretical discussion of the present results is undertaken with reference to Newton's original corpuscular theory of light. It is argued that his failure to predict that light travels more slowly in water than in air arose from the inadequacy of his mechanical theory rather than his assumptions about the elementary composition of light.

**I. Introduction**

Measurements of the speed of light in liquids and solids have had a decisive influence on the development of mechanical theories in science and in formulating models on which to visualize the fundamental processes of nature. The phenomenon of light refraction was already a subject of keen interest to the ancient scholars in Greece and Egypt, but it took many centuries of further study before it became clear that such effects are directly related to the fact that light travels with different speeds through air and water and other transparent materials. Two laws of refraction were discovered very early on, but it was not until the seventeenth century before the Dutch astronomer, Snell, was able to show that the sines of the angles of incidence and refraction always have a constant ratio for a given pair of media.

Experiments of this genre became the focus of a seminal argument about whether light in its elementary form is a particle or a wave. Newton concluded on the basis of his corpuscular theory of optical phenomena that particles of light travel faster in a dense medium such as water or glass than they do in air or free space. Belief in this theory was virtually abandoned a century and a half later when in 1850 Foucault was able to show that light actually travels more slowly in water than in air. The latter experiment was based on Fizeau's mechanical shutter method, which has also been the model for most subsequent measurements of the speed of light in dispersive media [1-3]. It involves the detection of interference between two light waves originating from the same source. The slower speed of light in dense media is explained by the fact that the wavelength of the radiation is decreased while the corresponding frequency remains unchanged. Little more than a decade later Maxwell formulated his electromagnetic theory and after another twenty years Hertz was able to confirm that it gave a correct description of the transmission of both visible light and radio waves of much lower frequency.

Yet Newton's theory of the particle nature of light received new impetus in 1905 through Einstein's interpretation of the photoelectric effect [4] and later from observations of collisions between x-rays and electrons in the Compton effect [5]. These experiments can only be successfully analyzed in terms of highly localized entities with a definite energy and momentum, later designated as photons by Lewis [6], which are very similar to the corpuscles of light envisioned by Newton.

The question thus arises whether it is possible to measure the speed of single photons without taking advantage of any of the wave properties of light such as interference. A fairer test of the particle hypothesis would be to accurately measure the elapsed time that it takes for a photon to travel a known distance from a light source to a suitable detector, much as one goes about determining the velocity of an ordinary object such as a train or a baseball. Recent advances in time-correlated single photon counting detection [7] open up an interesting possibility in this direction, as will be discussed in detail in the following section. On the basis of the present experimental investigation it has proven possible to measure the speed of light in water for three different wavelengths by timing the motion of single photons emitted from a laser source. The subsequent discussion of these results then considers the question of why Newtonian mechanics led to an erroneous prediction of the relative speeds of light in air and water some 300 years ago.

## II. Experimental Procedure

The technique employed to measure the speed of light in water in the present study has been implemented in past work to study relaxation effects in biological materials [8]. The underlying idea is to detect single photons over a period of time which have been used to irradiate a given substance. The method makes use of electronics which can measure the elapsed time between the firing of a laser pulse and the arrival of one of its photons at a photomultiplier tube (PMT) located some distance away. Before discussing exact details of the experimental procedure, a brief introduction to the model on which it is based will be given below.

*A. Statistics of Speed Measurements*

A simple way to visualize how the present experimental procedure enables a quantitative measurement of the speed of light in dispersive media is shown in the schematic diagram of Fig. 1. Analogy is made to the common procedure used to evaluate the results of a swimming race over a fixed distance AB. The basic idea is to start the clock at the moment the swimmer dives into the water and then to stop it immediately after the designated position at the end of the pool is reached. There are clearly two sources of error, corresponding to inaccuracies in initiating the timing at the proper moment and then later in stopping it precisely. In addition, one must be certain that the clock itself is functioning properly so that it gives an accurate value for the elapsed time to be measured. Because of the high speed of light, the sizes of the errors associated with the setting and stopping of the clock electronically are too large to allow the speed of any one photon to be determined within the desired level of accuracy. The present method overcomes this deficiency by relying on the fact that the errors in question are quite systematic and follow a definite statistical pattern.

If the race is judged by a large number of official timers, one can catalog their individual errors as $t_1(n)$ for the time it takes each of them to set their clock after the swimmer starts to dive and $t_3(n)$ for the corresponding time it takes to stop their clock after the final position has been reached. In the photon experiments under discussion it is certain that each $t_1$ and $t_3$ value will be positive, but this characteristic is not critical to the success of the overall determination. If the time actually required by the swimmer to complete the race in a fair manner is designated as $t_2$, then the total elapsed time t on a given clock n will be

$$t(n) = t_2 + t_3(n) - t_1(n). \qquad (1)$$

Without more specific knowledge of the individual $t_1(n)$ and $t_3(n)$ values, it is impossible to obtain an accurate measurement of the time $t_2$ from these results, but if the distribution of these errors is reproducible to a sufficient degree, it is possible to obtain an accurate comparison of the times $t_2^A$ and $t_2^B$ for two different swimmers. In other words, by sub-traction of the total clock times for these two races as determined by each of the judges, through systematic cancellation of errors one obtains

$$t^A(n) - t^B(n) = t_2^A - t_2^B \tag{2}$$

in all cases.

It is relatively easy to check how well the statistical distribution $t_3(n) - t_1(n)$, which will hereafter be referred to as the instrument response, is reproduced in different situations. One can simply compare results for different sample sizes pertaining to the *same* race after appropriate normalization. In the experimental procedure to be described below it will be seen that the range of $t_3(n)$ values is far larger than for $t_1(n)$ because the detection of a single photon at the PMT is understandably a more delicate operation than is involved in recognizing when the corresponding laser pulse has been fired. As discussed above, however, this detail is a minor consideration in comparison with the overall reproducibility of the instrument response in the present scheme.

Beyond this, it is necessary to calibrate the chronograph with respect to some known time interval. In the present work this is done by assuming that each photon travels through air with the same speed c. Modern-day measurements of the refractive index of air find a value of 1.00029 [9], which is sufficently close to unity for our purposes. Actually, as will be discussed in more detail in Sect. IV, the measured speed of light is always that of the group velocity $v_g = c/n_g$ rather than the phase velocity $v_p = c/n$ [10]. The group index of refraction $n_g$ is obtained from measurements of n at different light frequencies $\omega$ as [1,10]

$$n_g(\omega) = n(\omega) + \omega \frac{dn}{d\omega}. \tag{3}$$

In air $n_g$ differs from n by one part in 50000, so again, we have just taken the light speed in air to be equal to c (299 792 458 m/s).

After this calibration has been done, one can then obtain the speed of light in water by measuring the time difference $t_2^{H_2O} - t_2^{air}$ over a known path length. Because of the greater dispersion of light in water, however, the difference between the respective n and $n_g$ values is much larger than in air, particularly for higher frequencies. This raises the question of whether there is a corresponding increase in the range of velocities of the photons as they travel through the denser medium. In the present experiment this effect would manifest itself through a broadening of the instrument response in water *vis-a-vis* air, so this is an additional point of interest in considering these results.

*B. Details of the Experimental Arrangement*

In order to carry out the timing measurements for single photons as discussed above, the setup sketched in Fig. 2 has been employed.

A coherent Mira 900-F Ti:Sapphire laser was used as the light source for these experiments. Mode-locked operation in the femtosecond regime results in pulses of 120 fs nominal width, which can be considered as a δ function when compared to the instrument response (as defined above) and to the time interval $t_2^{H_2O} - t_2^{air}$. The temporal width was periodically checked with an APE *MINI* autocorrelator. The wavelength of operation was chosen at 810 nm, because this gives the highest intensity. This is an important consideration when tripling the frequency, as lower intensities of the fundamental do not reach the adequate threshold for frequency conversion. The Ti:Sapphire laser was pumped by a cw diode-pumped $Nd:YVO_4$ laser (Coherent Verdi 5W) emitting 5 watts of (continuous) 532 nm light.

The uncertainty principle establishes that the length of the pulse is inversely related to the line width of the pulse [11]

$$\Delta E \, \Delta t \geq \hbar \, . \qquad (4)$$

In the subpicosecond regime, the effect on the wavelength of the light pulse is relevant. Application of eq. (4) indicates that the true wavelength of the pulse is 810 ± 3 nm. From now on, when the wavelength is given, the value of the uncertainty will be implied.

The repetition rate at the exit of the Ti : Sapphire laser is 76 MHz. In these experiments, however, we are limited by the response of the photodiode that reads the presence of a laser pulse (see below). The photodiode has an upper limit of operation at 5-6 MHz. To solve this problem, a Coherent model 9200 pulse picker is used to eliminate 19 out of every 20 pulses. Therefore, the repetition rate for the light pulses used in the experiment is 3.8 MHz.

The laser beam (810 nm, 3.8 MHz, 120 fs pulses) can now be sent into the dispersive medium. Alternatively, it can be sent to an INRAD Ultrafast Harmonic Generator, model 5-050, where blue (405 nm, 250 fs) or ultraviolet light (270 nm, 350 fs) is generated using LBO or BBO crystals. Equation (4) indicates that the wavelengths of the visible and ultraviolet pulses are 405.0 ± 0.4 nm and 270.0 ± 0.1 nm, respectively.

Once the desired wavelength is chosen, the beam is steered to a cylinder containing the dispersive medium ($H_2O$ in all the experiments described here). The cylinder was built using a glass tube (14 mm outer diameter) with 90° cuts at both ends. A quartz window [thickness: 3.175 mm (nominally 1/8 in.)] was attached at each end. The quartz faces are placed perpendicular to the incoming laser beam in order to avoid changing the length of the beam path due to different refraction angles in air *vs* water. The inside length of the cylinder was measured at 0.9455 ± 0.0003 m. During the experiment, the cylinder was either empty (*i.e.* filled with air) or filled with deionized water (R > 18 MΩ). The difference in the time it takes for the photons to travel this distance in the two media was measured as described below.

Prior to entering the cylinder (see Fig. 2), a glass flat window was introduced in the path of the laser beam to deflect ~ 4 % of the beam into a photodiode (Thor Labs DET 210). The window is placed at different positions depending on the wavelength of the beam. Upon detection of this fraction of the pulse, the photodiode sends a signal to the electronics controller to indicate that time counting must be started. Using the swimming race analogy, this is the moment when the official timers start their chronographs.

After passing through the cylinder, the laser beam is reflected by two mirrors, effectively making two consecutive 90° turns. The two mirrors are mounted on a rigid platform attached to a sliding track. This track is aligned perfectly parallel to the path of the laser beam before reaching the first mirror and to the path of the beam after leaving the second mirror. This guarantees that upon sliding the platform along the track, the beam will still reach the detector, but the total length of the path can be shortened or lengthened at will (within the constraints of the track size). If the difference in path lengths is known, the timing instrument can be calibrated by using the equation $\Delta t = \Delta L/c$. In practice, it is not possible to accurately measure the position of the platform at intermediate positions in the track. However, the track length (equivalent to the difference in position of the platform at the beginning and end of the track) can be measured with submillimeter accuracy (and the positions are perfectly reproducible). In the experiments reported here, the track length is $0.2115 \pm 0.0003$ m, so $\Delta L$ is actually twice this difference, $0.4230 \pm 0.0006$ m.

A pinhole located shortly before the detector (diam. ~ 1-2 mm) ensures that the position of the beam is not disturbed by sliding the mirror assembly. The beam then reaches a quartz cuvette containing a particulate suspension (creamer in water) that scatters the laser beam into the detector, located at a 90° angle to the incident beam. The laser beam cannot be sent directly to the detector, as the PMT cannot withstand such an intense photon flux, hence the use of the suspension.

Detection is accomplished by a Hammamatsu R3809U-50 PMT controlled by E&G electronics. This electronics setup includes the following components: an EG&G TRUMP-8k-W3 multichannel analyzer card to interface the system to a PC; two EG&G model 9307 fast discriminators; an EG&G model 457 biased time-to-amplitude converter; an EG&G model DB463 delay generator; an EG&G model 4001C/4002D NIM rack and power supply; and an EG&G model 556 high voltage power supply. The assembly works as a time-correlated single photon counting detector. When the system receives the signal from the photodiode, the "chronograph" is started, *i.e.* time starts counting. When a photon is detected by the PMT, the chronograph is stopped. If no photon is detected by the PMT, the time counter is reset when the photodiode registers a new laser pulse. If a photon is detected by the PMT, the time interval t between starting pulse and the arrival of the stopping signal is measured and recorded. This is repeated for the duration of the experiment until a statistical distribution of the time it takes from starting pulse to detection (and clock stoppage) is measured (see Fig. 3). Unfortunately, this profile is not a $\delta$ function. It has a definite width, mainly due to the differences in the speed of travel of the electrons through the layers of the photomultiplier tube. However, the overall statistical profile is quite reproducible. We use these profiles to define the temporal events, as will be described in more detail in Sect. III. This way of measuring time, between start and stop signals, requires that only one photon per laser pulse be registered. As a result, the detector shuts down upon receiving one photon, and resets when a new laser pulse is detected by the photodiode. This requires that the photon flux be rather small. If it were not, the statistical profile would be biased towards photons arriving at the PMT shortly after the initiation pulse. In practice, this means that the power supply (high voltage) for the PMT is set so that an upper limit of one photon per every 200 laser pulses is detected, *i.e.* a maximum of 20000 photons per second are recorded.

The dynamic range of the electronics comprises 8192 data points. The actual range can be varied. In these experiments it is set up to cover approximately 10 ns, so that each data point has a width of about 1.2 fs. We have found that the nominal dynamic range is not accurate enough for these measurements. Therefore, calibration (using the sliding track, see Fig. 2) is required to accurately measure $t^{H_2O}(n) - t^{air}(n)$. An alternative way of calibrating the time scale was used in some experiments (using only $\lambda = 270$ nm). In these experiments the steering mirrors were mounted directly on the laser table, so that the photons always traveled along a path of the same size. Calibration was accomplished by performing experiments using two different lengths of coaxial cable (RG-58A/U) to connect the left to the right side of the delay generator. For each medium (air and water), two sets of data were collected, with connecting cables differing in length by $0.3105 \pm 0.0003$ m. As the electronic signals travel at a speed of $0.66c$ [12], the difference in the x-axis position of the two sets of data is $1.569 \times 10^{-9}$ s. This number was used to calibrate the pixel temporal size in this set of experiments.

### III. Results of the Measurements

As mentioned in Sect. II.A, the analysis of the photon timings consists of three distinct comparisons. First, it needs to be shown that the instrument response is sufficiently reproducible to obtain a quantitative measurement of elapsed times. A simple test of this nature consists of a comparison of the distributions of the photon timings obtained under identical conditions but over different detection periods. An example of this type is given in Fig. 3 for light traveling through air with a wavelength $\lambda = 270$ nm. The corresponding distributions over time are brought to maximum overlap by multiplying the values obtained in the shorter period by a factor of 1.92. The difference (residuals) of the two normalized distributions is also shown in Fig. 3, from which it can be seen that the largest discrepancy is 365 counts in a given time slice (pixel), as compared to a total count at peak maximum of 10500. The locations of the two peak maxima are found to be the same.

In all, four such comparisons have been made under a variety of conditions and the largest discrepancy between values at the same location between the normalized distributions was found to be 957 counts (compared to 10500 counts at peak maximum). The location of the peak maximum differs by 15 pixel in this comparison, which corresponds to a time difference of ca. 20 ps (see below). In the other two cases the corresponding differences were 4 and 6 pixel, respectively. The full widths of the peaks at half maximum (FWHM) fall uniformly in the 100 ps range. In each case there is a fairly sudden rise in counting, but after the count maximum is reached there is always a characteristic shoulder in the distribution before counts cease to be recorded. It should be emphasized that the broadness of the peaks is due entirely to the instrument response, referred to as $t_3(n) - t_1(n)$ in Sect. II.A, as all the detected photons have traveled the same distance through air before reaching the PMT. Furthermore, as will be seen below, the shapes of the photon distributions are basically unchanged when the path length is varied or a different transparent medium is introduced along the path.

*A. Time-scale Calibration*

The next step in the experimental procedure is to compare the timing results obtained when light traverses two different tracks in air whose path lengths differ by 0.4230 m. The time required for the light to travel the latter distance is 1411 ps. The data in Fig. 4 demonstrate that almost identical distribution patterns are found in the two cases. The two peaks are brought to maximum overlap by a displacement of 1085 pixel (for $\lambda = 270$ nm). After normalization to 10500 counts at peak maximum, the maximal difference between respective photon counts over the entire range is 382. In general, these measurements indicate that such deviations are quite similar to what is observed when peaks corresponding to different timing durations for the *same* track are compared. The residuals curve given below the left-hand peak is a particularly good means of demonstrating this similarity (Fig. 4).

Analogous results for these two tracks were also obtained for light of 405 and 810 nm wavelengths. The maximum discrepancies in the normalized distributions (ca. 10500 counts at peak maxima) are 770 and 932 counts, respectively. The displacements required to bring the corresponding peaks into maximum overlap are 1076 and 1086 pixel, in good agreement with the value mentioned above for 270 nm light. In addition, an analogous series of runs over the same two path lengths was made with water in the cylinder instead of air. At 270 nm maximal overlap of the distributions is obtained with a shift of only 1051 pixel, but for 405 (see Fig. 5) and 810 nm the corresponding shifts are in much closer agreement with the above results obtained with air in the cylinder (1085 and 1086 pixel, respectively). Inspection of the six shifts, 1051, 1076, 1085, 1085, 1086, and 1086 pixel shows that the first data point is suspect. Application of the Q test [13] indicates that this point should be discarded from further analysis. The average value of the remaining five shifts is 1084 ± 6 pixel. On this basis, one obtains a ratio for the time calibration of 1.302 ± 0.010 ps/pixel (1084 pixel = 1411 ps).

*B. Light Speed Measurements in Water*

We are now in a position to compare the photon times of flight (TOF) with and without water in the cylinder. An example of the corresponding photon count distributions is given in Fig. 6 (long track and $\lambda$ = 405 nm). Again it can be seen that the shapes of these profiles are quite similar (see residuals plotted under the first peak). That obtained with water in the cylinder must be shifted ahead by 911 pixel to obtain maximal overlap with the corresponding distribution obtained when the cylinder is filled with air. The maximal discrepancy in the respective counts is 489, again compared to a value at peak maximum of about 10000. A second measurement of this type has been carried out with the short track, in which case a shift of 899 pixel is found to give the optimal coincidence.

The average of these two shifts is 905 ± 6 pixel, which according to the above calibration, corresponds to a time delay of 1178 ps in water relative to air. The time for light to traverse the cylinder (0.9455 m) in air is known to be 3154 ps, giving a ratio of $v_{air} / v_{H_2O}$ of 1.374 ± 0.006. This value lies close to the group index of refraction $n_g$ at this wavelength of light (Fig. 7), which is inferred [see eq. (3)] from available measured n values (between 760.82 and 396.8468 nm [14]), namely 1.3790. The error is calculated by taking into account the accumulated errors of measuring the pixel size, track length, cell size, and the error in the determination of the position of the peak in air *vs* in water.

A second measurement of the water-air TOF difference has been made for light of $\lambda$ = 810 nm. This wavelength lies slightly to the red of the above values for which n values are available (Fig. 7) [14], but the corresponding $n_g$ result can still be accurately estimated by extrapolation (1.3423). A comparison of the measured photon count distributions with the cylinder filled with water and air (long track), respectively, is given in Fig. 8, along with a plot of the residuals. After normalization of the two peaks, maximal overlap occurs for a shift of 841 pixel, with a maximal discrepancy of 418 counts (peak maximum of 10100) over the entire range. The corresponding shift for the short track is 840 pixel. In this case the maximum discrepancy is relatively large (1102 counts), after normalization to 10900 counts and optimal displacement. The average of the peak shifts represents a time delay of 1094 ps, corresponding to a $v_{air} / v_{H_2O}$ ratio of 1.347 ± 0.006. This result thus lies higher than the above $n_g$ value obtained from refractive index data, whereas the measured $v_{air} / v_{H_2O}$ ratio at 405 nm is slightly lower than its corresponding $n_g$ value (Fig. 7). Taken together these results indicate that the speed of the single photons is $c/n_g$ in each case and that the experimental error is not of a particularly systematic nature.

Finally, a third determination has been made at 270 nm. This wavelength lies too far in the uv to be able to give an accurate value for $n_g$ based on the available refractive index data (Fig. 7). At $\lambda$ = 397 nm the measured n value is 1.3435, while $n_g$ can be estimated to be 1.381. The present measured photon count distributions (long track) are given in Fig. 9 for the cases with and without water in the cylinder. After shifting and normalization, the maximum discrepancy over the peak region is 852 counts, 8.5 % of the value at peak maximum. The corresponding shift is 1103 pixel. The shift for the short track comparison is 1137 pixel, so the discrepancy between these two values is larger than for the other two wavelengths. From the average of 1120 ± 17 pixel one obtains a value for the $v_{air} / v_{H_2O}$ ratio of 1.463 ± 0.010. The experiment has also been carried out employing a different electronics arrangement (see Sect. II.B), with nearly the same result (1.461). The fact that the spread in the above peak shifts is somewhat larger than for the other two cases employing longer laser wavelengths seems consistent with the fact that one is faced with additional experimental difficulties this far in the uv region. Alignment of the beam is difficult as it cannot be seen with the naked eye, and the intensity is quite low (of the order of several tens of nW) so that it does not register very easily on fluorescing paper. Furthermore, the low intensity also requires that very high voltages are used to power the PMT (2900 – 3000 V *vs* 2200 – 2400 V for 405 and 810 nm). The higher voltages result in lower signal-to-noise ratios and in the presence of additional features (which are quite reproducible) in the instrument response (see, for instance, the shoulder to the right of each peak present in Figs. 3, 4, and 9).

## IV. Newtonian Mechanics and Light Speed

The experimental data discussed above can be interpreted in a straightforward manner as a series of repetitive trials in which a single photon of a given laser pulse travels a definite path under identical conditions before it is detected by a photomultiplier tube. The electronics employed to obtain the elapsed time of each photon's journey along this path are not capable of giving an accurate determination of its velocity in a single trial, but the distribution of flight times resulting from a large number of such measurements follows a definite pattern which is reproducible to a high degree. The shape of the instrument response exhibits only minor variations for different paths traversed by the light, independent of the length of the track or the media through which it passes. The photons emitted from the laser source have a very narrow range of velocities close to c, and it would appear from the present experiments that all that happens when they pass through water is that they are all decelerated by the same amount.

Such an interpretation is clearly very much in line with Newton's seventeenth century views on the elemental composition of light, and yet the measured change in velocity stands in direct contradiction to his prediction that the light speed should be greater in water than in air. It is therefore of interest to examine more closely the line of reasoning which led to this incorrect conclusion. His arguments were based primarily on observations of the refraction of light in dispersive media (see Fig. 10). Because light is always bent more toward the normal when it enters water from air (Snell's Law), it is necessary to assume according to Newton's mechanical theory that there is an attractive force in the medium of higher n which causes the particles of light to be accelerated there. This conclusion was reached before the pioneering experiments of the late nineteenth century which led to quantum mechanics and special relativity, however, so it is interesting to consider what information these theories provide which was not known to Newton.

As is discussed in more detail in a companion article [15], there is good reason to believe that the potential acting on the photons is more attractive in water than in air, just as Newton said. Instead, it was his method of computing the velocity of the photons from this fact which is faulty. First of all, one must distinguish carefully between velocity and momentum in this case, because it is far from certain that the inertial mass of the photons is the same in both media. The fact that their potential energy is lower in water while their total energy E is unchanged implies that their kinetic energy is greater than in air. It is important to note that the conclusion that the momentum p of the photons also increases is strongly supported by the quantum mechanical relation,

$$p = \hbar k = h/\lambda. \qquad (5)$$

It is well known that the wavelength of light is inversely proportional to the index of refraction, so it follows from eq. (5) that the photon momentum must be greater in water, consistent with Newton's assumption of an attractive force acting in this medium.

The problem with his argumentation arises because it is assumed that the velocity of the photons must also be greater because their inertial mass does not change as they pass from one medium to another. The correct result for the photon velocity, as verified by the present time-correlated single photon counting detection experiments, is obtained from Hamilton's canonical equations of motion as

$$v = dE / dp, \tag{6}$$

also as discussed in a companion paper [15]. Although this is an expression from classsical mechanics [16], it would not have been of any use to Newton because he had no way of evaluating the above derivative. Use of Planck's relation,

$$E = \hbar \omega, \tag{7}$$

in conjunction with eq. (5) overcomes this difficulty, however, leading to the observed result:

$$v = v_g = \frac{d\omega}{dk} = c/n_g, \tag{8}$$

that is, that the velocity of single photons is equal to the group velocity as defined above via eq. (3). In the case of water this expression gives a value for the speed of light which is less than that in air, even though eq. (5) on which it is based clearly indicates that the opposite ordering holds for their momenta in these two media.

It is thus easy to understand why Newton was led to his erroneous prediction for the speed of light in dispersive media. Without the benefit of the quantum mechanical relations of eqs. (5,7) and, to a lesser extent, the evidence for the variation of mass with potential energy from the theory of special relativity, it was impossible for him to know that the existence of an attractive force within a given medium does not always imply an increase in particle velocity. Nonetheless, it is interesting to note that this failure does not constitute proof that his corpuscular theory of light is inoperable, rather only that the mechanical theory he used to arrive at his velocity prediction is inadequate for this purpose.

**V. Conclusion**

The present study has employed a novel method for determining the velocity of light in dispersive media which is based on time-correlated single photon counting. Advantage is taken of the statistical regularity in the time required to send an electronic signal from a photomultiplier tube to a chronograph. A characteristic instrument response is observed when measuring the times of flight of single photons traveling a fixed distance through air. As a result, it is possible to obtain an accurate calibration for the chronograph by recording the shift in the location of the instrument response when the distance moved by the photons between source and detector is varied. This procedure allows photon TOF differences to be determined to an accuracy of ca. 10 ps.

By inserting a glass cylinder approximately 1.0 m in length along the path of the photons, it is then possible with this apparatus to determine the ratio of the velocity of light in water to that in air. A key observation is that the shape of the instrument response is very nearly the same whether the above cylinder is filled with air or water (see Figs. 3–6, 8, 9). It is thus a quite straightforward matter to measure the corresponding TOF difference by noting the shift required to bring the two counting distributions to maximum coincidence and employing the above calibration.

Measurements have been carried out for light of three different wavelengths. For 405 and 810 nm the photon velocities are found to be in good agreement with the corresponding group velocity ($c/n_g$) results deduced from refractive index measurements. For the shorter of these two wavelengths an $n_g$ value of $1.374 \pm 0.006$ is obtained, which is 0.005 smaller than that inferred from the available n values, whereas at 810 nm, a result of $1.347 \pm 0.006$ is found, which is too high by 0.005 based on a slight extrapolation of the corresponding n values in the neighborhood of this wavelength. Two determinations have also been made for $\lambda = 270$ nm light. They indicate an $n_g$ value of $1.463 \pm 0.010$, but there is insufficient refractive index data available so far in the uv region to allow for a meaningful comparison in this case.

From a theoretical point of view the most interesting result of the present investigation is that the shape of the instrument response for single photons in a laser beam appears to be totally unaffected by their passage through a dispersive medium. This seems to imply that the velocity distribution for photons corresponding to a given wavelength of light is a d function in all media and therefore does not contribute to the width of the counting profile attributed to the instrument response in Figs. 3–6, 8, 9. If the velocity profile had a width, i.e. if photons from light of a given wavelength in air were capable of propagating with different velocities, then one would expect that this distribution would broaden significantly when the light enters a medium of much higher refractive index such as water. The method employed does not require interference of two light waves emanating from the same source, in contrast to the classical measurements of Bergstrand, Michelson and Houston [1-3] using variations of the Fizeau mechanical shutter technique, or to the more recent conjugate photon experiments of Steinberg *et al.* [17], which make use of an extension of the Hong-Ou-Mandel interferometer [18]. Photons are simply sent one at a time from the laser to a PMT and clocked in a manner which is analogous to what occurs in a conventional swimming competition. The clear indication is that all photons subjected to the same conditions (wavelength of light, track size and nature of the media through which they pass) travel at exactly the same speed. In particular, they simply undergo uniform deceleration in passing from air into water.

The present experiments are thus supportive of a particle theory of light which has much in common with the the views professed by Newton in the late seventeenth century. Accordingly, the momentum of each photon is given by the quantum mechanical relations of eqs. (5-7) to be $p = n\hbar\omega/c$, and thus is greater in water than in air. The corresponding velocity is $v = dE/dp$, however, which, again with the help of quantum mechanics, is the group velocity of light, $v_g = d\omega / dk = c / n_g$. Newton's erroneous prediction of a higher light speed in water than in air can thus be traced to deficiencies in his mechanical theory at that point in time rather than to a fundamental misunderstanding of the elemental composition of light. These matters are discussed in more detail in a companion article [15], but in the last analysis the best way to settle the longstanding argument of whether light consists of particles or waves is to measure the momentum transferred to electrons or nuclei as a result of radiative processes occurring in dispersive media. In the absence of such new experiments, however, it would appear from the present study that a theoretical analysis in terms of single photons traveling the distance between source and detector at the same well-defined velocity for a given dispersive medium and wavelength of light is perfectly consistent with all available measurements.

**Acknowledgments**

The authors wish to thank Mr. James R. Hodgson for providing his glassblowing services which were essential for the present study. The instrumentation used in the present experiments was purchaced with funds from National Science Foundation grant CHE-9709034 (PLM). This work was also supported in part by the Deutsche Forschungsgemeinschaft within the Schwerpunkt Programm "Theorie relativisticscher Effekte in der Chemie und Physik schwerer Elemente." The financial support of the Fonds der Chemischen Industrie is also hereby gratefully acknowledged.

**References**

1. E. Bergstrand, Arkiv Physik **8**, 457 (1954).

2. A. A. Michelson, Rep. Brit. Assoc. Montreal 1884, p. 56.

3. R. A. Houston, Proc. Roy. Soc. Edinburgh A **62**, 1 (1944).

4. A. Einstein, Ann. Physik **17**, 132 (1905).

5. A. H. Compton, Phys. Rev. **21**, 715 (1923); **22**, 409 (1923).

6. G. N. Lewis, Nature **118**, 874 (1926).

7. X. S. Xie and R. C. Dunn, Science **265**, 361 (1994).

8. P. J. Reid, D. A. Higgins and P. F. Barbara, J. Phys. Chem. **100**, 3892 (1996).

9. G. Shortley and D. Williams, Elements of Physics (Prentice-Hall, Englewood Cliffs, N.J., 1961), p. 528.

10. L. Brillouin, Wave Propagation and Group Velocity (Academic Press, New York, 1960), pp. 1-3.

11. L. I. Schiff, Quantum Mechanics (McGraw-Hill, New York, 1955), p. 7.

12. D. DeMaw, The Radio Amateur's Handbook (American Radio League, 1968), p. 336.

13. D.P. Shoemaker, C.W. Garland and J.W. Nibler, Experiments in Physical Chemistry (5$^{th}$ Edition, McGraw-Hill, New York, 1989), p.35.

14. H. Niedrig, Bergmann Schaefer Lehrbuch der Experimentalphysik, Vol. 3, Optik (de Gruyter, Berlin, 1993), pp. 209-210.

15. R. J. Buenker, P. L. Muiño and A. M. Thompson, Khim. Fys. **23**, No. 2, 111 (2004).

16. H. Goldstein, Classical Mechanics (Addison-Wesley, Reading, Mass., 1959), p. 227.

17. A. M. Steinberg, P. G. Kwiat and R. Y. Chiao, Phys. Rev. Lett. **68**, 2421 (1992); **71**, 708 (1993).

18. C. K. Hong, Z. Y. Ou and L. Mandel, Phys. Rev. Lett. **59**, 2044 (1987).

**Figure Captions**

Fig. 1. Schematic diagram showing the three time intevals which are involved in the electronic clocking of a racing event: $t_1$ for starting the clock after the object has left the starting gate A, $t_2$ for the actual travel time from A to B, and $t_3$ for stopping the clock after arrival of the object at B. The total elapsed time registered on the clock is thus $t = t_2 + t_3 - t_1$.

Fig. 2. Experimental setup used for measuring the various time intervals needed to obtain the velocity of light in water for three different wavelengths of light.

Fig. 3. Distribution of photon counts as a function of pixel time slice for 270 nm light with the test cylinder of Fig. 2 filled with air. Two sets of results are shown, corresponding to different durations of the counting. The data shown on the baseline are the residuals obtained by subtracting these two distributions after appropriate normalization to give maximum coincidence.

Fig. 4. Distribution of photon counts as a function of pixel time slice for 270 nm light with the test cylinder filled with air. The two peaks correspond to different path lengths within the apparatus ($\Delta L = 0.4230$ m). The residuals curve below the left-hand peak is obtained by appropriate normalization and shifting to bring both peaks into maximum coincidence. The magnitude of the shift is used to calibrate the chronograph.

Fig. 5. Distribution of photon counts as a function of pixel time slice for 405 nm light with the test cylinder filled with $H_2O$. The two peaks correspond to different path lengths within the apparatus ($\Delta L = 0.4230$ m). The residuals curve below the left-hand peak is obtained by appropriate normalization and shifting to bring both peaks into maximum coincidence. The magnitude of the shift is used to calibrate the chronograph.

Fig. 6. Distribution of photon counts as a function of pixel time slice for 405 nm light with the test cylinder filled alternately with air (left-hand peak) and $H_2O$ (right-hand peak) and using the long track (see Fig. 2). The residuals curve below the left-hand peak is obtained by appropriate normalization and shifting to bring both peaks into maximum coincidence. The magnitude of the shift is used to obtain the difference in elapsed times for light to travel through the cylinder ($\Delta L$ = 0.9455 m) in the two cases and hence the ratio of the velocities of light in air and water.

Fig. 7. Variation of the group index of refraction $n_g$ of water with the wavelength of light as obtained from a polynomial fit of experimental refractive indices n [13] and using eq. (3). Comparison is made with the present measured results for the ratio of $v_{air}$ to $v_{water}$ at 810, 405 and 270 nm.

Fig. 8. Distribution of photon counts as a function of pixel time slice for 810 nm light with the test cylinder filled alternately with air (left-hand peak) and $H_2O$ (right-hand peak) and using the long track (see Fig. 2). The residuals curve below the left-hand peak is obtained by appropriate normalization and shifting to bring both peaks into maximum coincidence. The magnitude of the shift is used to obtain the difference in elapsed times for light to travel through the cylinder ($\Delta L$ = 0.9455 m) in the two cases and hence the ratio of the velocities of light in air and water.

Fig. 9. Distribution of photon counts as a function of pixel time slice for 270 nm light with the test cylinder filled alternately with air (left-hand peak) and $H_2O$ (right-hand peak) and using the long track (see Fig. 2). The residuals curve below the left-hand peak is obtained by appropriate normalization and shifting to bring both peaks into maximum coincidence. The magnitude of the shift is used to obtain the difference in elapsed times for light to travel through the cylinder ($\Delta L$ = 0.9455 m) in the two cases and hence the ratio of the velocities of light in air and water.

Fig. 10. Schematic diagram showing the refraction of light at an interface between air and water. The fact that the light is always bent more toward the normal (Snell's Law) led Newton to believe that there is an attractive potential in the denser medium which causes the particles of light to be accelerated.

Fig.1

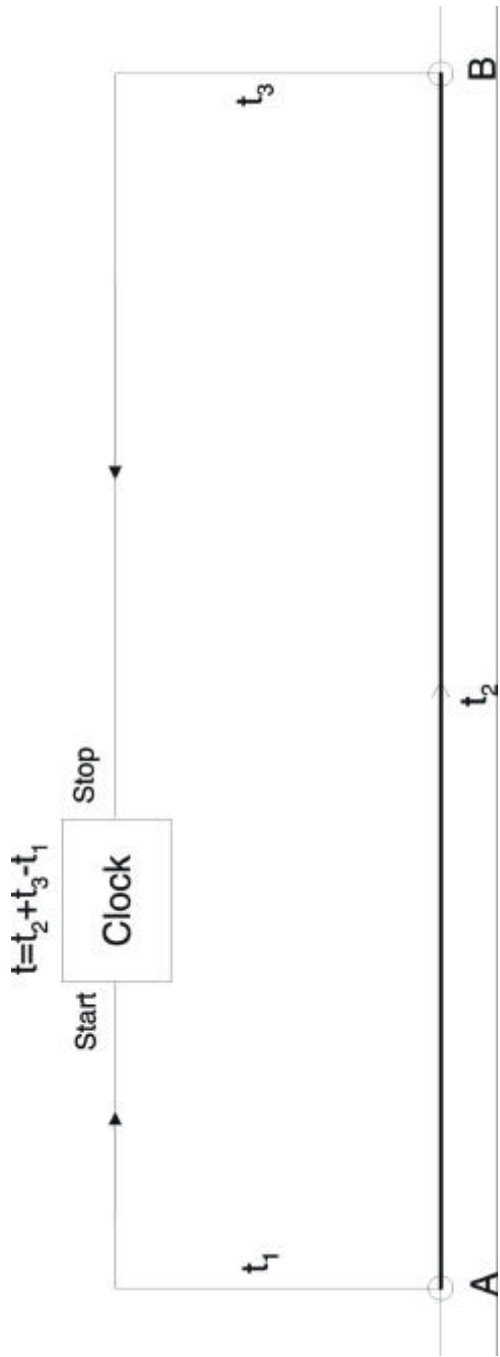

Fig.2

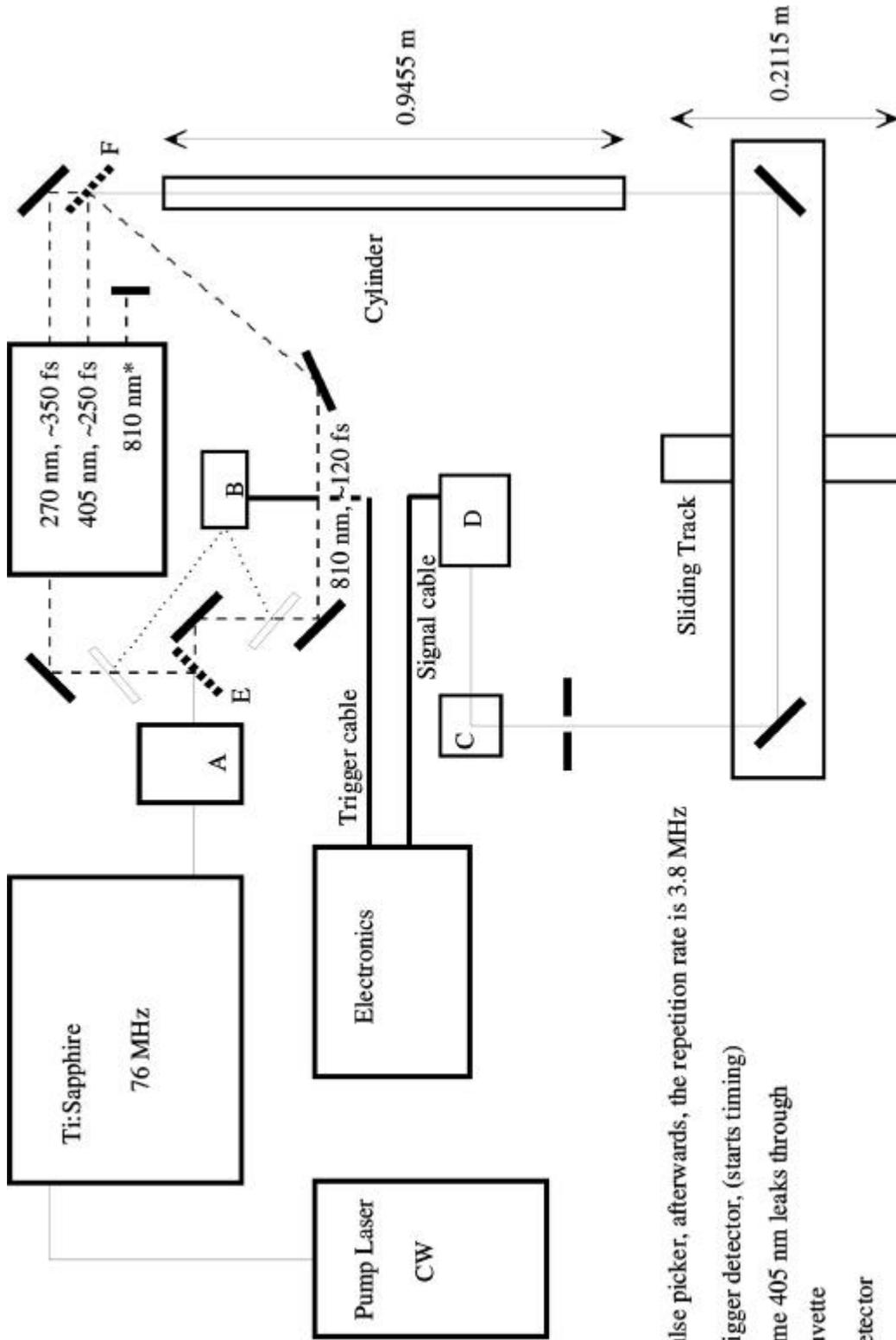

A: Pulse picker, afterwards, the repetition rate is 3.8 MHz

B: Trigger detector, (starts timing)

*: Some 405 nm leaks through

C: Cuvette

D: Detector

E: Mirror removed for 810 nm experiments

F: Mirror removed for 270 nm experiments

Note: Discontinuous lines indicate that the object may be removed in a specific experiment, (or that the laser beam is not present in some of the experiments).

Fig.3

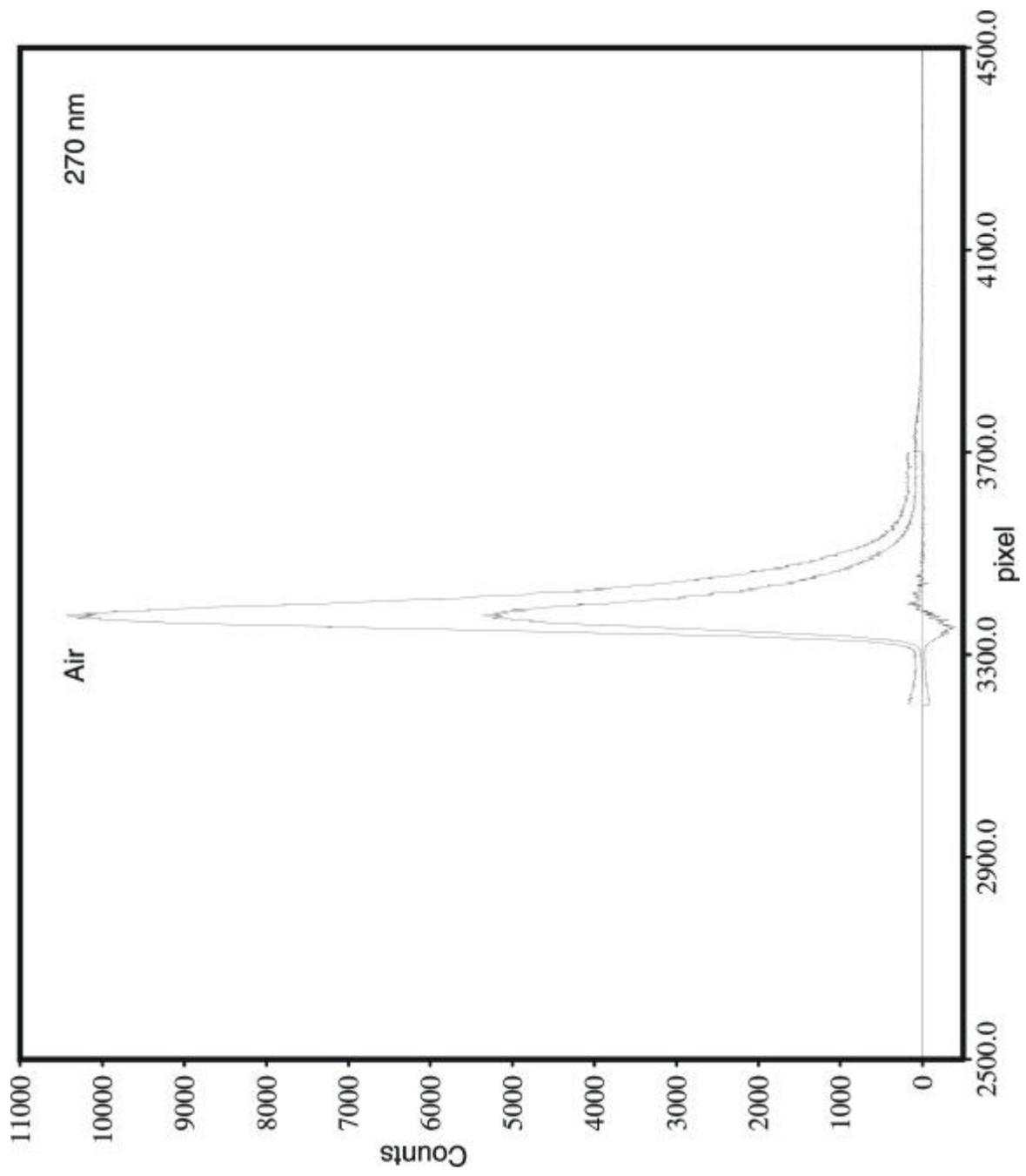

Fig.4

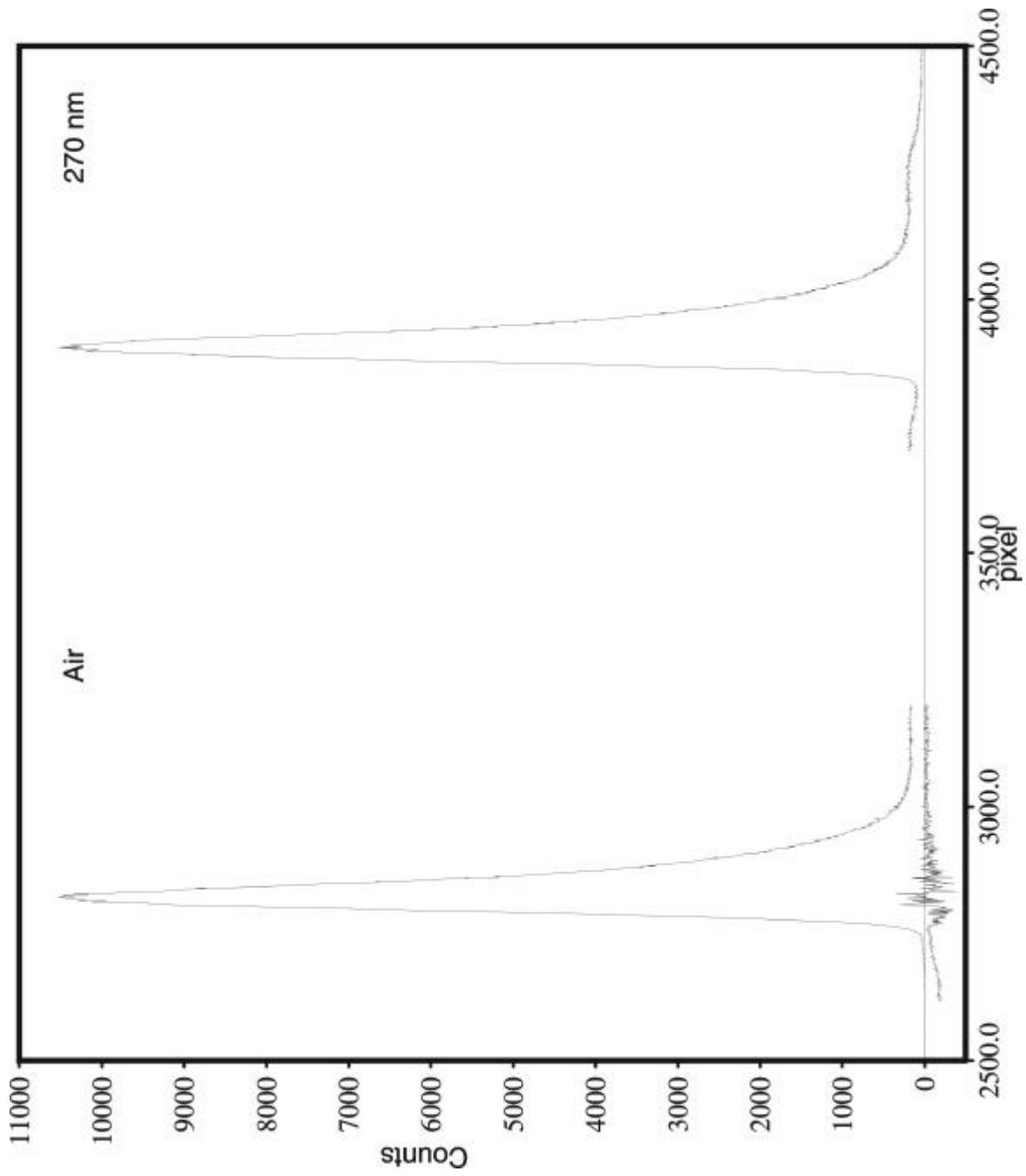

Fig.5

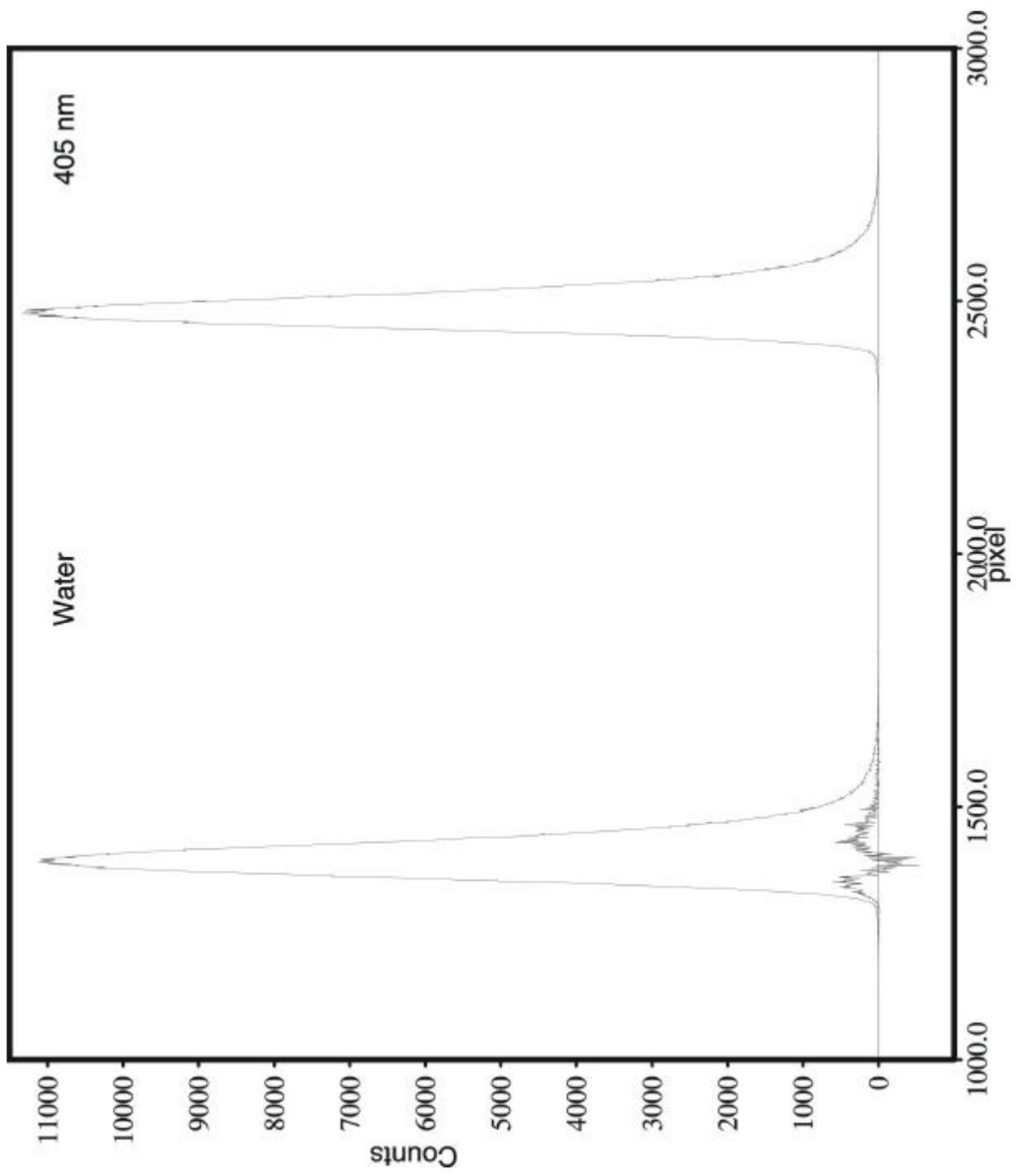

Fig.6

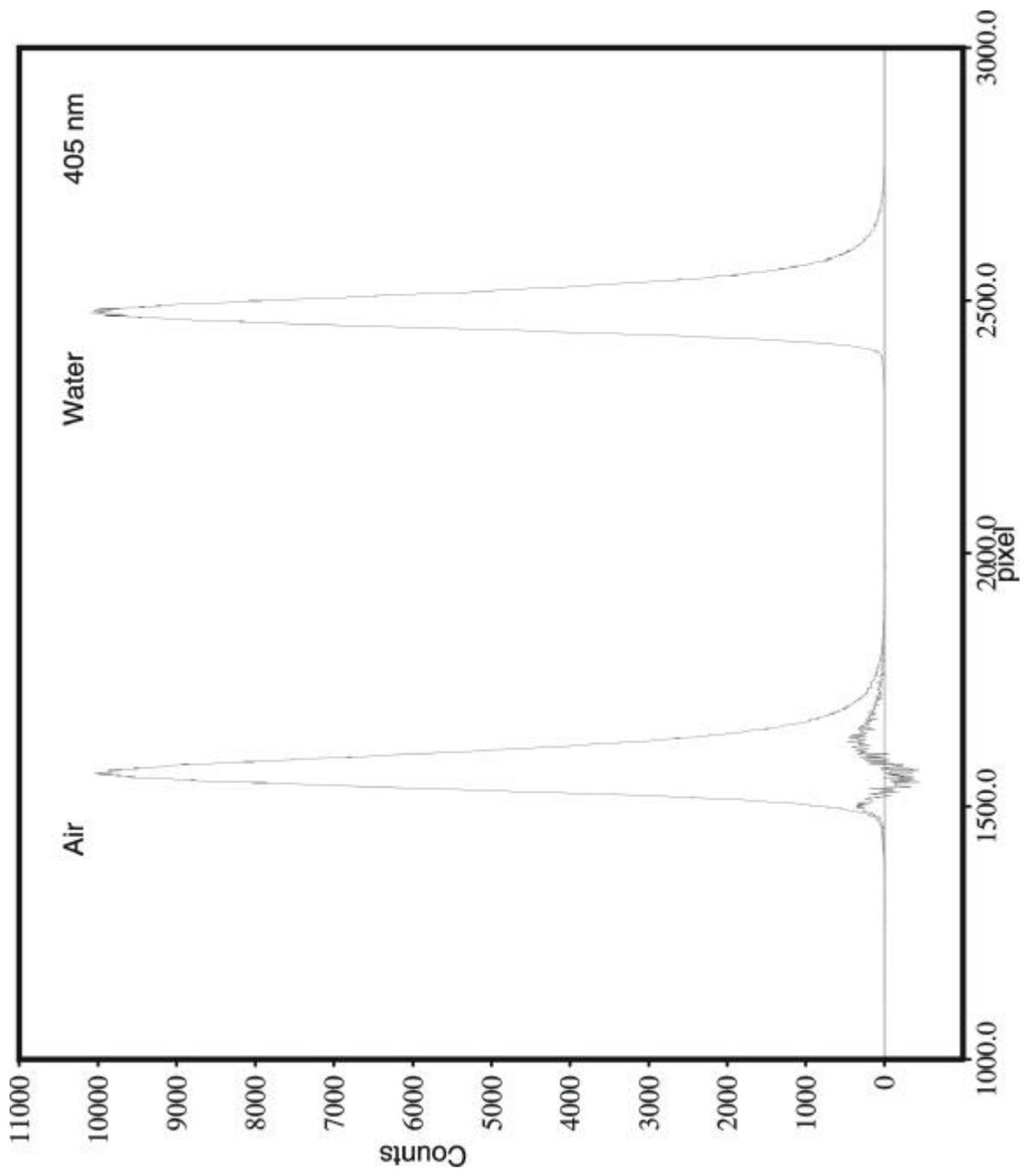

Fig.7

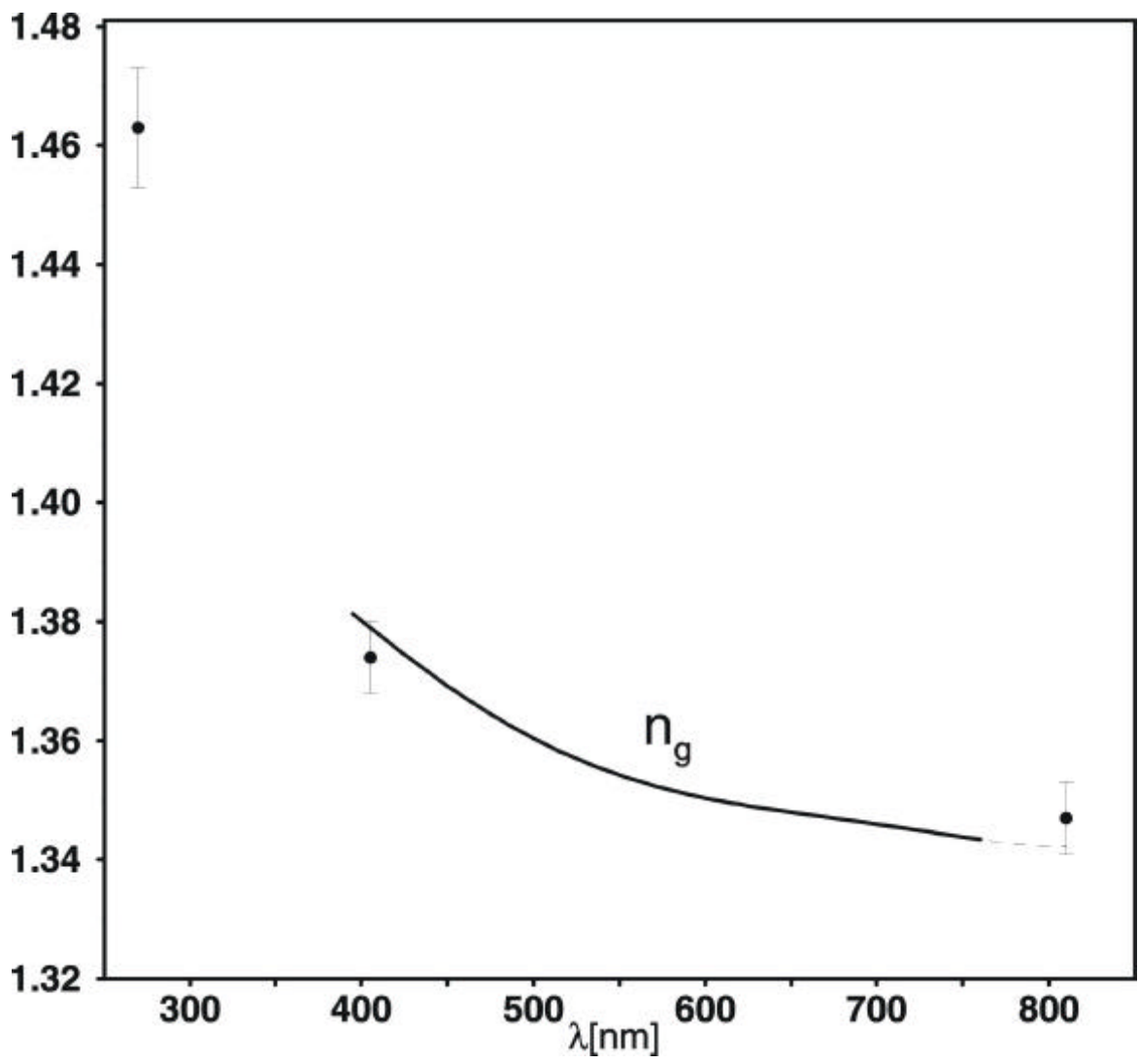

Fig.8

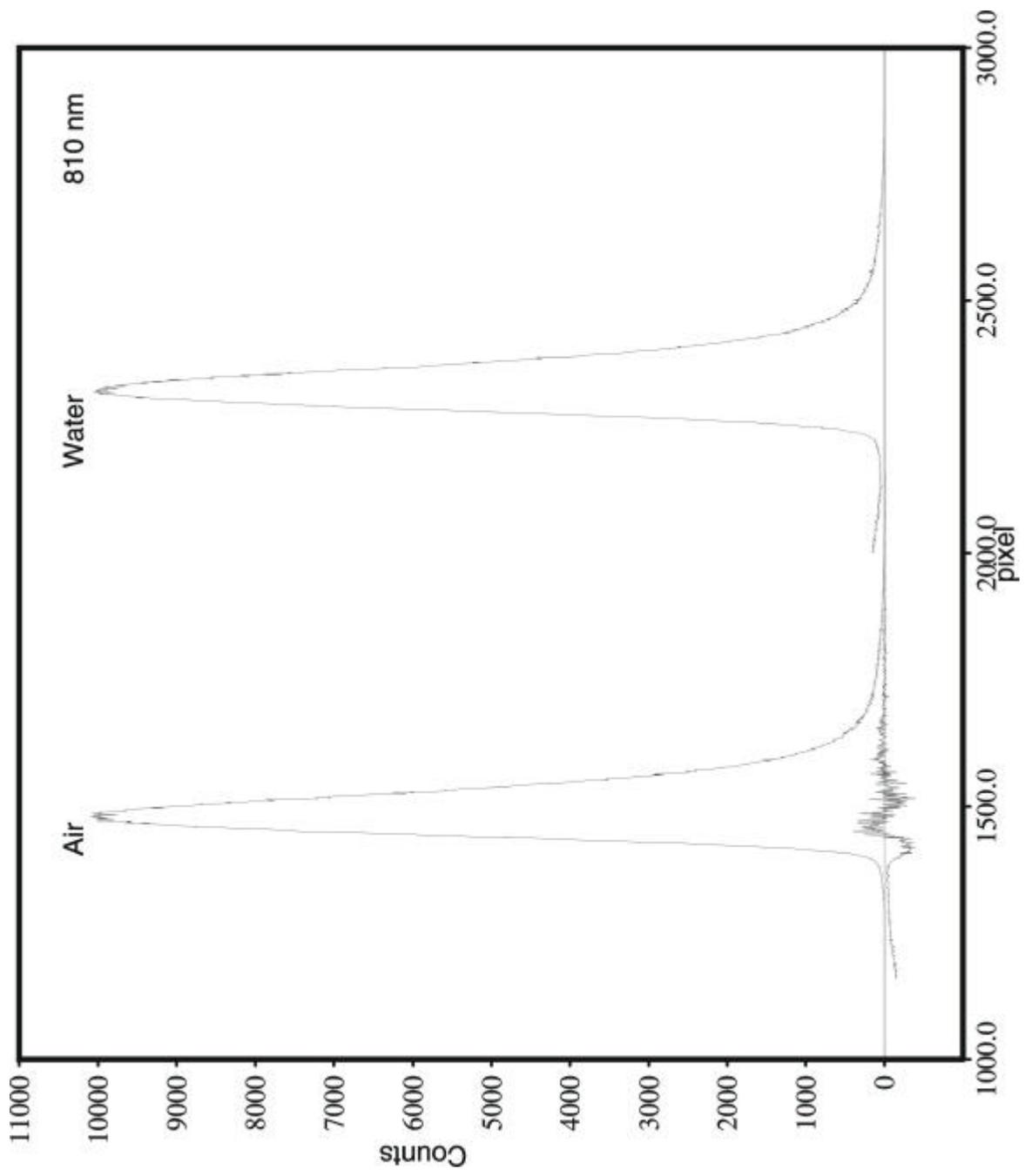

Fig.9

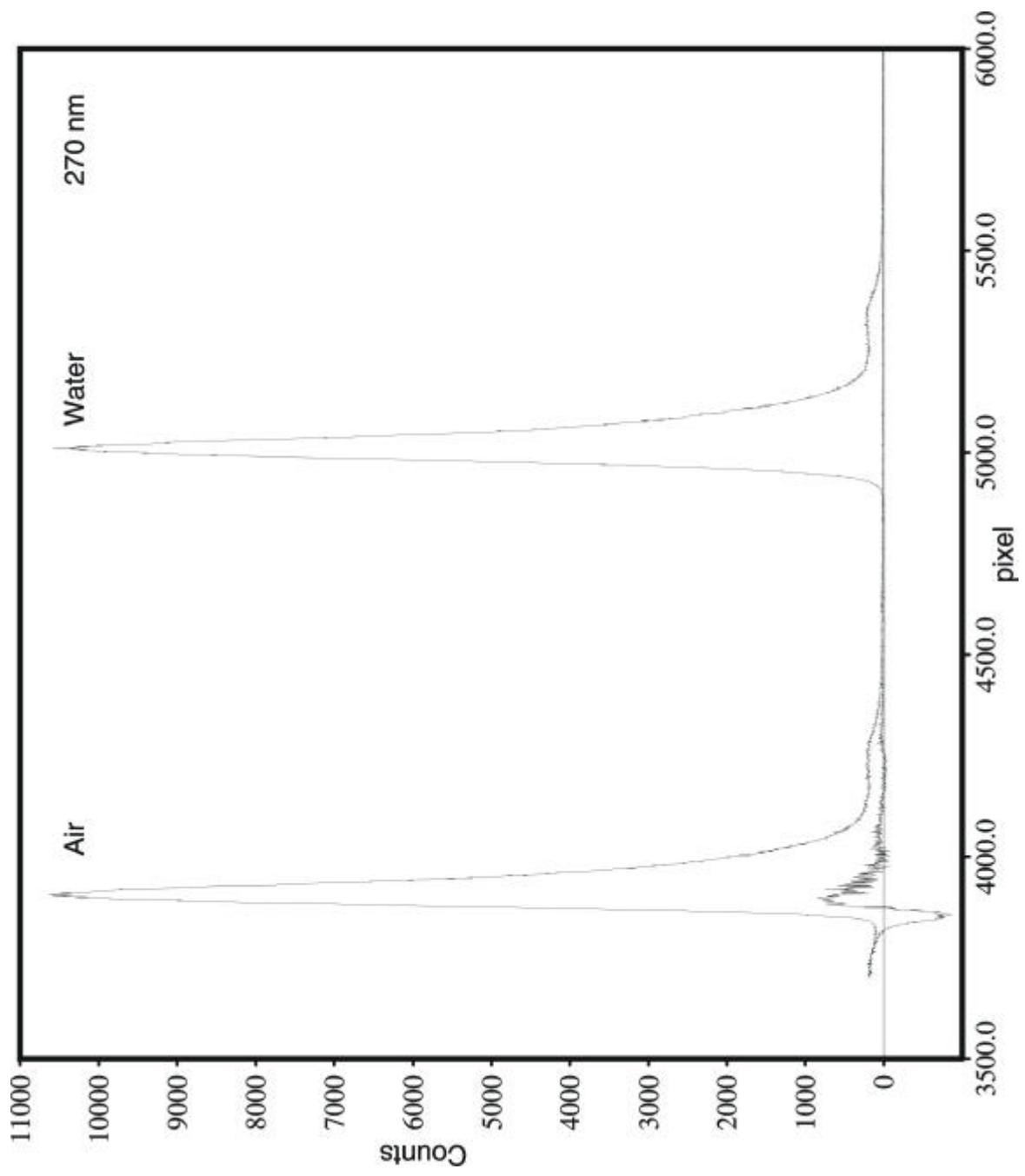

Fig.10

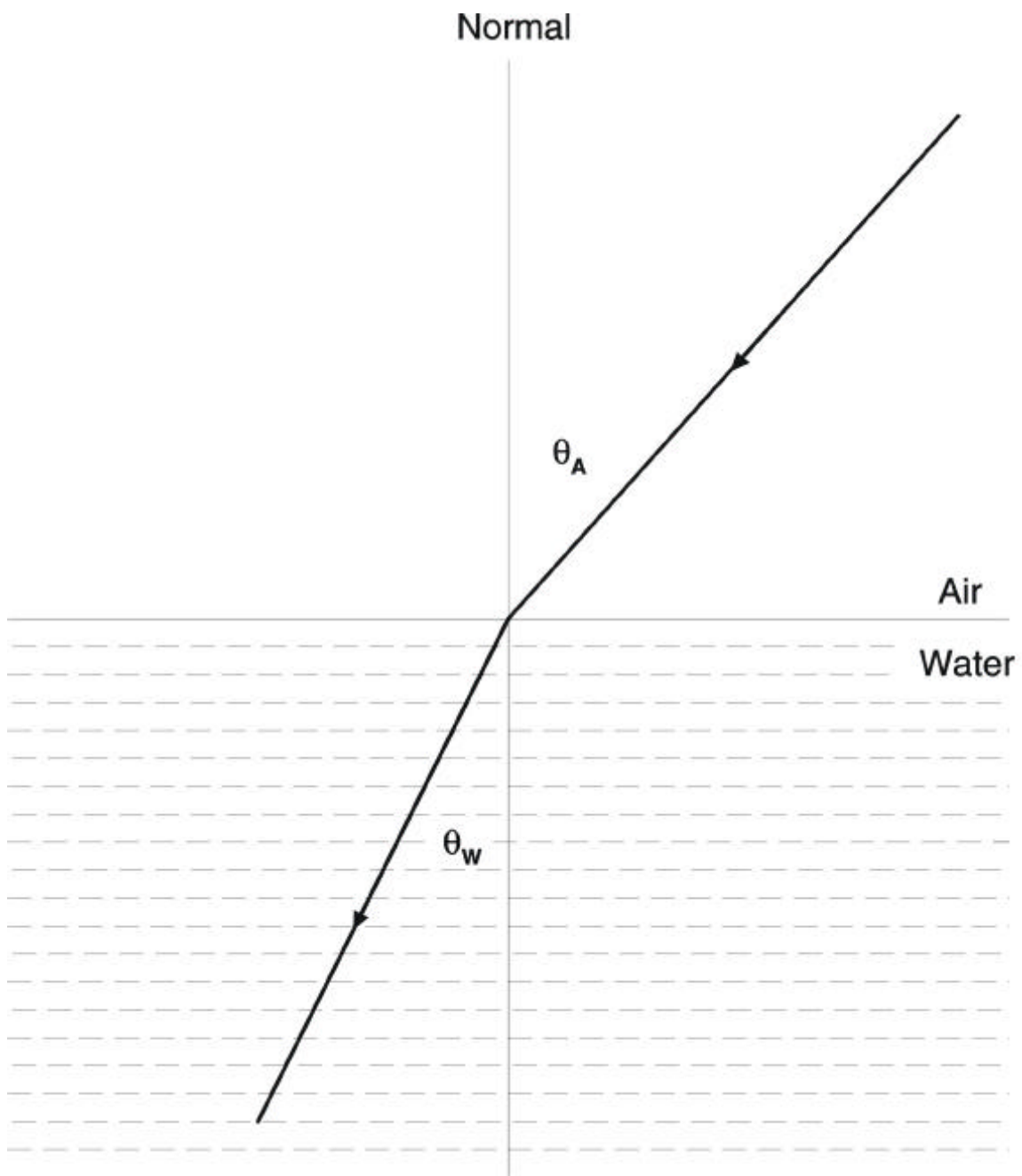